\documentclass[a4paper,11pt]{article}
\usepackage{pos-icrc}

\usepackage[utf8]{inputenc}
\usepackage[T1]{fontenc}
\usepackage{graphicx}
\usepackage{enumitem}
\usepackage{natbib}
\usepackage{gensymb}
\usepackage{amsmath}
\usepackage{amssymb}
\usepackage{mathtools}
\usepackage{slashed}
\usepackage{aps_macros}

\title{ EAS optical Cherenkov signatures of tau neutrinos for space and suborbital detectors }
 \ShortTitle{EAS optical Cherenkov signals of $\nu_\tau$ for space and suborbital detectors}

\author*[a]{Mary Hall Reno}
\author[b]{Tonia M. Venters}
\author[c,d]{John F Krizmanic}
\forColl{POEMMA and JEM-EUSO}

\affiliation[a]{University of Iowa,\\
  Department of Physics and Astronomy, Iowa City, USA}
\affiliation[b]{
NASA/Goddard Space Flight Center, Code 661\\
  8800 Greenbelt Rd, Greenbelt, Maryland 20771 USA}

\affiliation[c]{University of Maryland, Baltimore County, Center for Space Sciences and Technology\\
  1000 Hilltop Cir, Baltimore, Maryland 21250 USA}

\affiliation[d]{Center for Research and Exploration in Space Science \& Technology
\\
NASA/Goddard Space Flight Center, Code 661\\
  8800 Greenbelt Rd, Greenbelt, Maryland 20771 USA \hfill\vskip 0.1in}

\emailAdd{mary-hall-reno@uiowa.edu}
\emailAdd{tonia.m.venters@nasa.gov}
\emailAdd{john.f.krizmanic@nasa.gov}

\abstract{Multi-messenger observations of transient astrophysical sources have the potential to characterize the highest energy accelerators and the most extreme 
environments in the Universe. Detection of neutrinos, in particular tau neutrinos generated by neutrino oscillations in transit from their sources to Earth, is possible for neutrino energies above 10 PeV using optical Cherenkov detectors imaging upward-moving extensive air showers (EAS). These EAS are produced from Earth-interacting tau neutrinos leading to tau leptons that subsequently decay in the atmosphere. We compare neutrino detection sensitivities for generic short- and long-burst transient neutrino sources and sensitivities to the diffuse neutrino flux for the second generation Extreme Universe Space Observatory on a Super-Pressure Balloon (EUSO-SPB2) balloon-borne mission and the proposed space-based Probe of Extreme Multi-Messenger Astrophysics (POEMMA) mission.}

\FullConference{37$^{\rm{th}}$ International Cosmic Ray Conference (ICRC 2021)\\
		July 12th -- 23rd, 2021\\
		Online -- Berlin, Germany}

\begin{document}
\maketitle

\section{Introduction}
Transient astrophysical sources such as tidal disruption events, binary neutron star mergers and the merger of black hole pairs are expected to generate signals in gravitational waves, cosmic rays, photons and in neutrinos. Over cosmic distance scales, the flavor ratio of two muon neutrinos to one electron neutrino will oscillate to approximately equal numbers of all three flavors of neutrinos (and anti-neutrinos). At high energies, signals of tau neutrino fluxes may be observed through tau neutrino induced up-going air showers that produce an optical Cherenkov signal, the focus of this work. Using the Earth as a tau neutrino converter, taus are produced in $\nu_\tau$ charged current interactions \cite{Fargion:2003kn}. Taus may exit directly or proceed through a series of decays and neutrino reinteractions called regeneration. Exiting taus that decay into a shower in the atmosphere can produce optical and radio Cherenkov signals, and at the highest energies, air fluorescence. 

\begin{figure}
      \vspace{-3.5cm}
          \includegraphics[width=0.7\textwidth,angle=270]{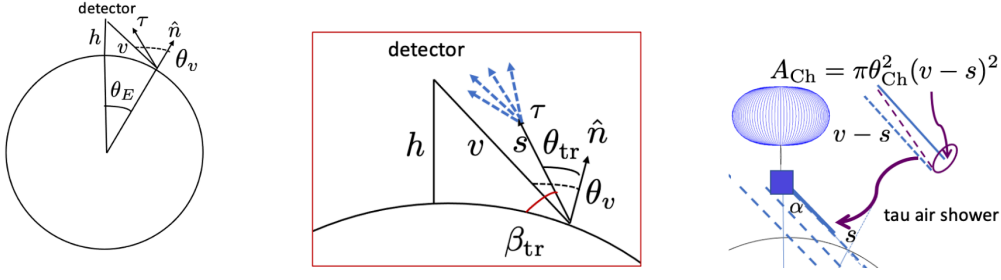}
       \vspace{-3cm}
       \caption{Definition of angles and distances for diffuse flux and point source fluence sensitivities.
       }
       \label{fig:geo}
     \end{figure}

The optical Cherenkov signal of up-going tau air showers is the target of the two missions describe here:
the Probe of Extreme MultiMessenger Astrophysics (POEMMA) \cite{Olinto:2020oky} with a pair of satellites proposed for a 2028 launch, and the
second generation Extreme Universe Space Observatory on a Super-Pressure Balloon (EUSO-SPB2) \cite{Eser} pathfinder telescope scheduled to launch in May 2023. Both missions also have a fluorescence telescope as well, discussed elsewhere in these proceedings.
Parameters of these satellite and sub-orbital telescopes are listed in table \ref{tab:telescopes}. With reference to fig. \ref{fig:geo}, table \ref{tab:telescopes}
shows the altitude $h$, the angular coverage below the limb $\Delta\alpha$ and the corresponding maximum Earth emergence angle (relative to the surface plane)
$\beta_{\rm tr}$, the azimuthal coverage $\Delta \phi$ and the minimum
photon number density at the respective detectors $\rho_\gamma^{\rm min}$ consistent with a reduction of the air glow background to less than 1\%
 \cite{Reno:2019jtr}. POEMMA has two observing modes: stereo with both telescopes pointing to the same spot on the ground, and dual, where the telescopes point to two different locations. Stereo mode has the advantage of a lower photon number density threshold, most important at lower energies, while the dual mode increases the observing area by a factor of 2. Telescopes can slew over time frames of order 500 s \cite{Venters:2019xwi}.

\begin{table}[htp]
\begin{center}
\begin{tabular}{|c|c|c|}
\hline
Mission & POEMMA  Stereo (Dual) & EUSO-SPB2 \\
\hline
\hline
$h$ & 525 km & 33 km\\
\hline
$\Delta \alpha$ & $7^\circ$ & $6.4^\circ$\\
\hline
$\Delta \phi $ & $30^\circ$ & $12.8^\circ$ \\
\hline
$\rho_\gamma^{\rm min}$  &  20 (40) $\gamma$/m$^2$ & 200 $\gamma$/m$^2$
\\ \hline
$\beta_{\rm tr}^{\rm max}$ & 19.6$^\circ$ & 10.8$^\circ$\\ \hline
\end{tabular}
\end{center}
\caption{Altitude $h$, angular coverage below the limb $\Delta\alpha$, azimuthal coverage $\Delta \phi$, minimum photon number density on the telescope surface $\rho _\gamma^{\rm min}$ and maximum
Earth emergence angle $\beta_{\rm tr}^{\rm max}$, }
\label{tab:telescopes}
\end{table}%

POEMMA's altitude $h=525$ km is a clear advantage over the EUSO-SPB2 sub-orbital altitude in coverage of the Earth's surface for detection of the diffuse astrophysical neutrino flux, a factor of $\sim 70$ enhanced geometric effective aperture of the Earth. For point source detection, POEMMA's advantage is not in effective area, but instead that it has full sky coverage over a year, during which time the orbital plane precesses through almost seven periods. In the next section, we compare the diffuse neutrino flux sensitivities of POEMMA and EUSO-SPB2. The target of opportunity (ToO) sensitivities are compared in the following section. We conclude with a short discussion of some of the modeling uncertainties.
For the range of neutrino energies relevant to these detectors, above $10^7$ GeV, neutrinos and antineutrinos have equal cross sections, so in our discussion, ``neutrino'' refers to both particle and antiparticle.

\section{Diffuse neutrino flux}

The sensitivity to the diffuse neutrino flux for a detector with effective aperture
$\langle A\Omega\rangle$ and  observing time including duty cycle $f_{\rm obs}$
for three neutrino flavors is 
\begin{equation}
F_{\rm sens} = \frac{2.44\times 3}{\ln (10)\times E_{\nu_\tau}\times \langle A\Omega
\rangle \times t_{\rm obs}}\,,
\end{equation}
assuming no background, at a 90\% confidence level over a decade of energy.
With the variables defined in fig. \ref{fig:geo}, a detector's effective aperture is given by \cite{Motloch:2013kva}
\begin{equation}\label{eq:apereqn}
\left<A\Omega\right> \left(E_{\nu_{\tau}}\right)= \int_{S}\int_{\Delta\Omega_{\rm tr}} P_{\rm obs}\ \hat{r}\cdot\hat{n}\, dS \, d\Omega_{\rm tr}\ ,
\end{equation}
where 
the observation probability
$P_{\rm obs}$ depends on the probability $p_{\rm exit}$ for $\nu_\tau \to \tau$ with an exiting $\tau$, the probability for the tau to decay in the atmosphere $p_{\rm decay}$, and the detection probability $p_{\rm det}$:
\begin{align}\label{eqnpobs}
P_{\rm obs} &= \int p_{\rm exit}\left(E_{\tau}|E_{\nu_{\tau}},\beta_{\rm tr}\right) \nonumber \\
& \times \left[\int ds'\, {p}_{\rm decay}(s') p_{\rm det}\left(E_{\tau},\theta_{E},\beta_{\rm tr},s'
\right)\right]\, dE_{\tau}\, ,
\end{align}
A full discussion of the diffuse neutrino flux sensitivity can be found in ref. \cite{Reno:2019jtr}.

\begin{figure}
       \includegraphics[width=.48\textwidth]{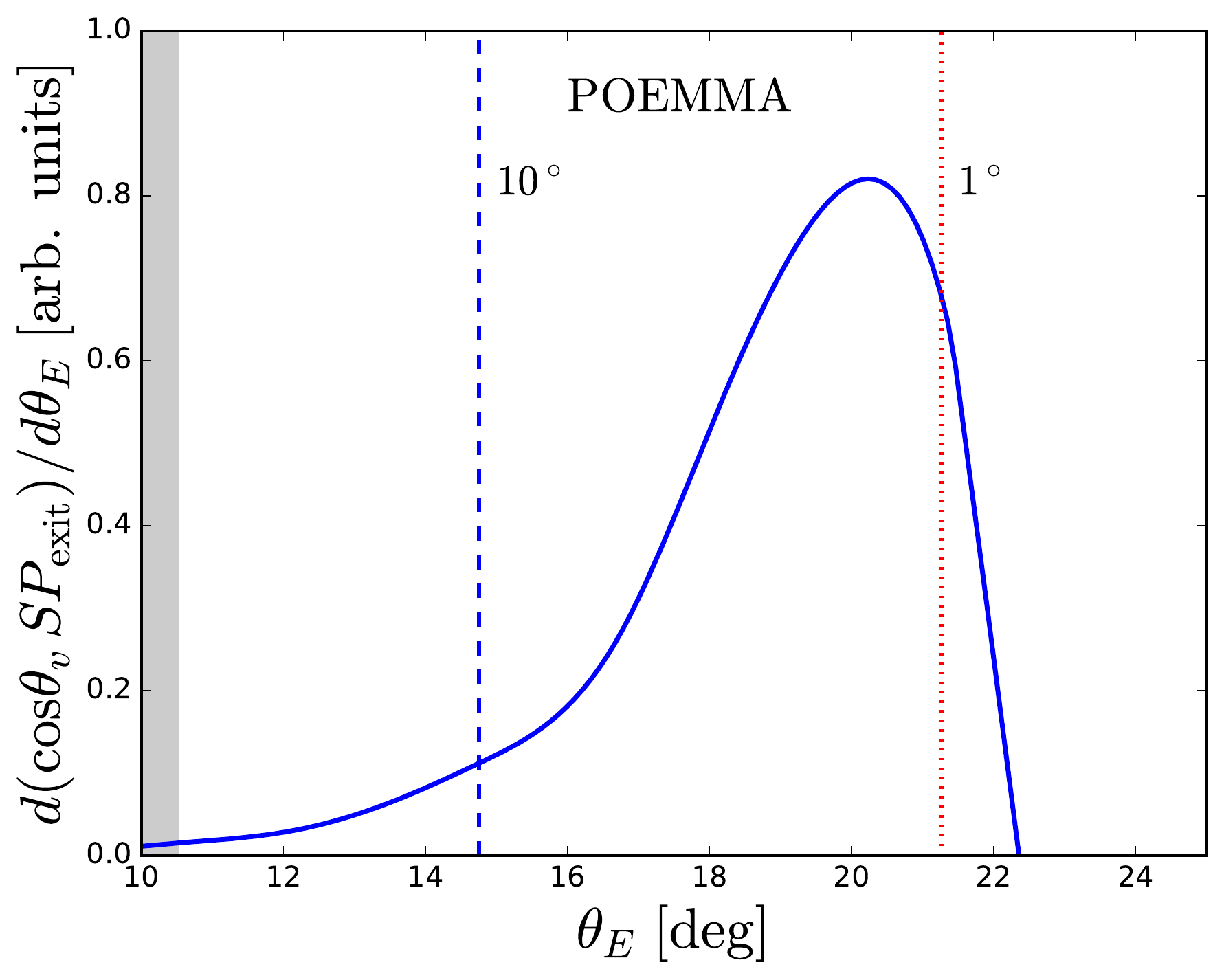}
       \includegraphics[width=.48\textwidth]{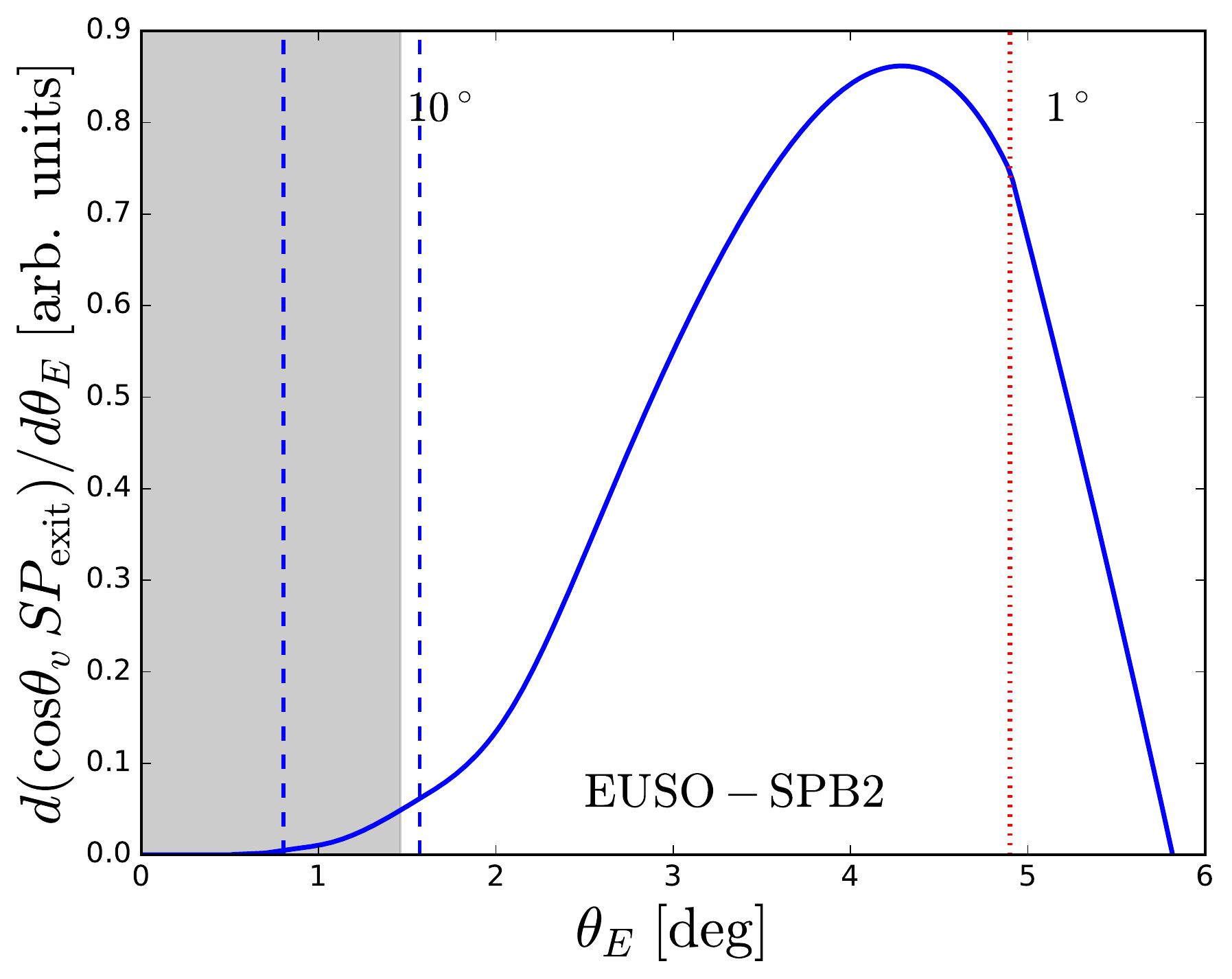}       
       \caption{
      Comparison of differential effective aperture for POEMMA and EUSO-SPB2 for $E_\nu=10^9$ GeV, as a function of $\theta_E$. The dotted vertical lines are labeled with the corresponding value of Earth emergence angle $\beta_{\rm tr}$.
       }
       \label{fig:pexit-geo}
     \end{figure}

Figure \ref{fig:pexit-geo} shows part of  the integrand in eq. (\ref{eq:apereqn}): the product of the geometric portion of the effective area $\hat{r}\cdot \hat{n} dS=\cos\theta_v dS$ with the tau exit probability $p_{\rm exit}$ for $E_\nu=10^9$ GeV, accounting for the angular ranges $\Delta \alpha$ and $\Delta\phi$ for POEMMA and EUSO-SPB2 shown in table \ref{tab:telescopes}.
The shaded regions are outside of the respective $\Delta \alpha$ ranges. For reference, the Earth emergence angles $\beta_{\rm tr}=1^\circ,\ 10^\circ$ are shown by vertical lines. Both telescopes cover the most important range of $\theta_E$ ($\beta_{\rm tr}$).

\begin{figure}
\hskip 1.5in
       \includegraphics[width=.48\textwidth]{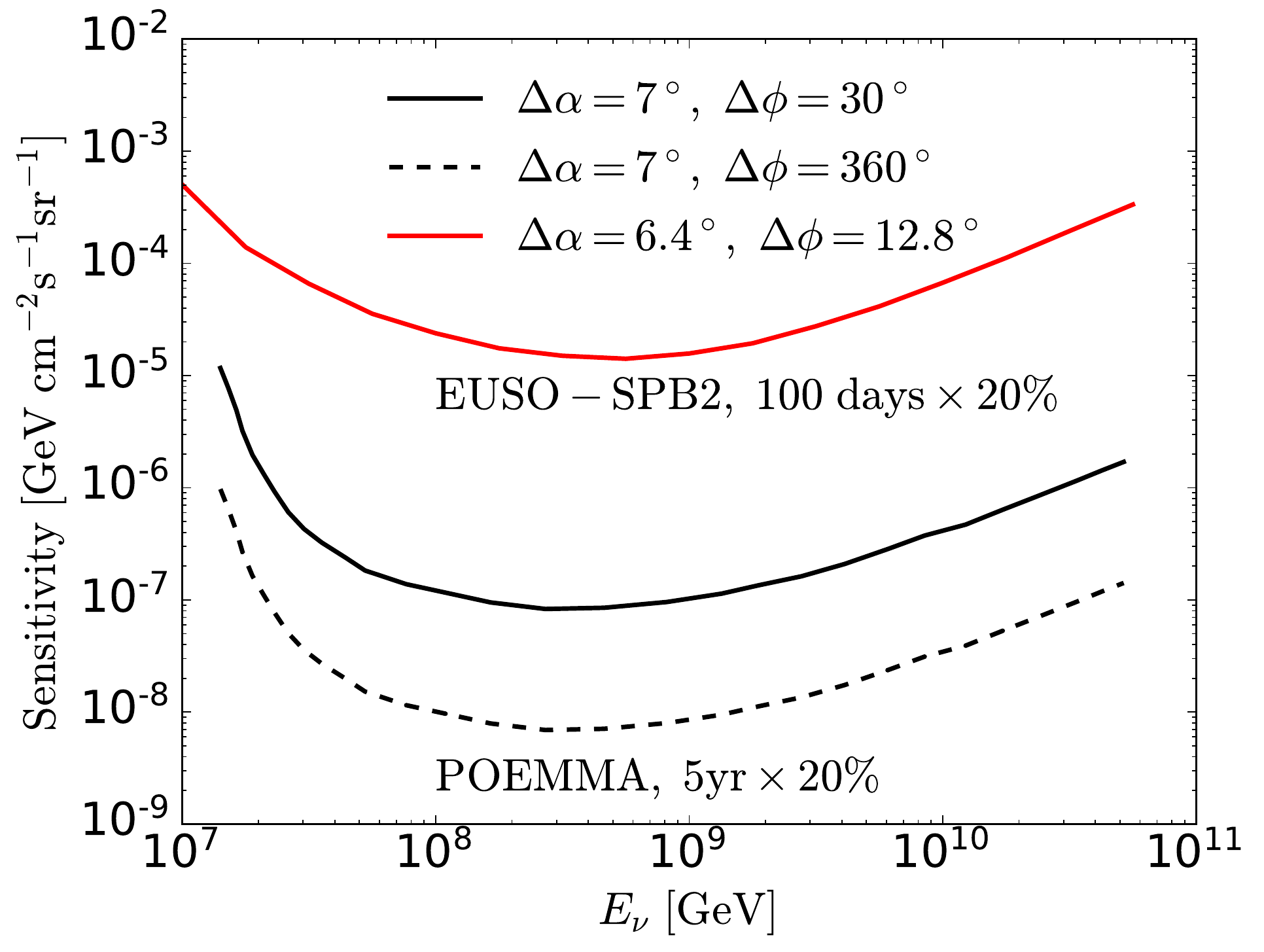}
\caption{The sensitivity of POEMMA and EUSO-SPB2 to the all-flavor diffuse neutrino flux.}
\label{fig:diffuse-limits}
\end{figure}

The sensitivity of both instruments to Cherenkov signals of up-going air showers degrades at low energies because $p_{\rm exit}$ is smaller for low energies than high energies. Tau decays dominate tau propagation, and the decay length of the tau in vacuum is
$500$ m$\times (E_\tau/10^7$ GeV).
Figure \ref{fig:diffuse-limits} shows the diffuse all-flavor neutrino flux sensitivity of POEMMA, EUSO-SPB2 and the sensitivity of a $\Delta\phi=360^\circ$ version of POEMMA \cite{Reno:2019jtr}. Without $\Delta\phi=360^\circ$, POEMMA will not be competitive with other instruments. The 5-year planned running time, a factor of $\sim 18$ times longer than EUSO-SPB2, and the advantage in geometric effective aperture account
are only partly offset by the enhanced detectability of EUSO-SPB2 that comes from being closer to the showers as they develop in the atmosphere. Overall, POEMMA with $\Delta\phi=30^\circ$ is a factor of $\sim 200$ more sensitive than EUSO-SPB2 to the diffuse astrophysical neutrino flux over most of the neutrino energy range.

\section{Target of opportunity neutrino fluence}

\begin{figure}
       \includegraphics[width=.48\textwidth]{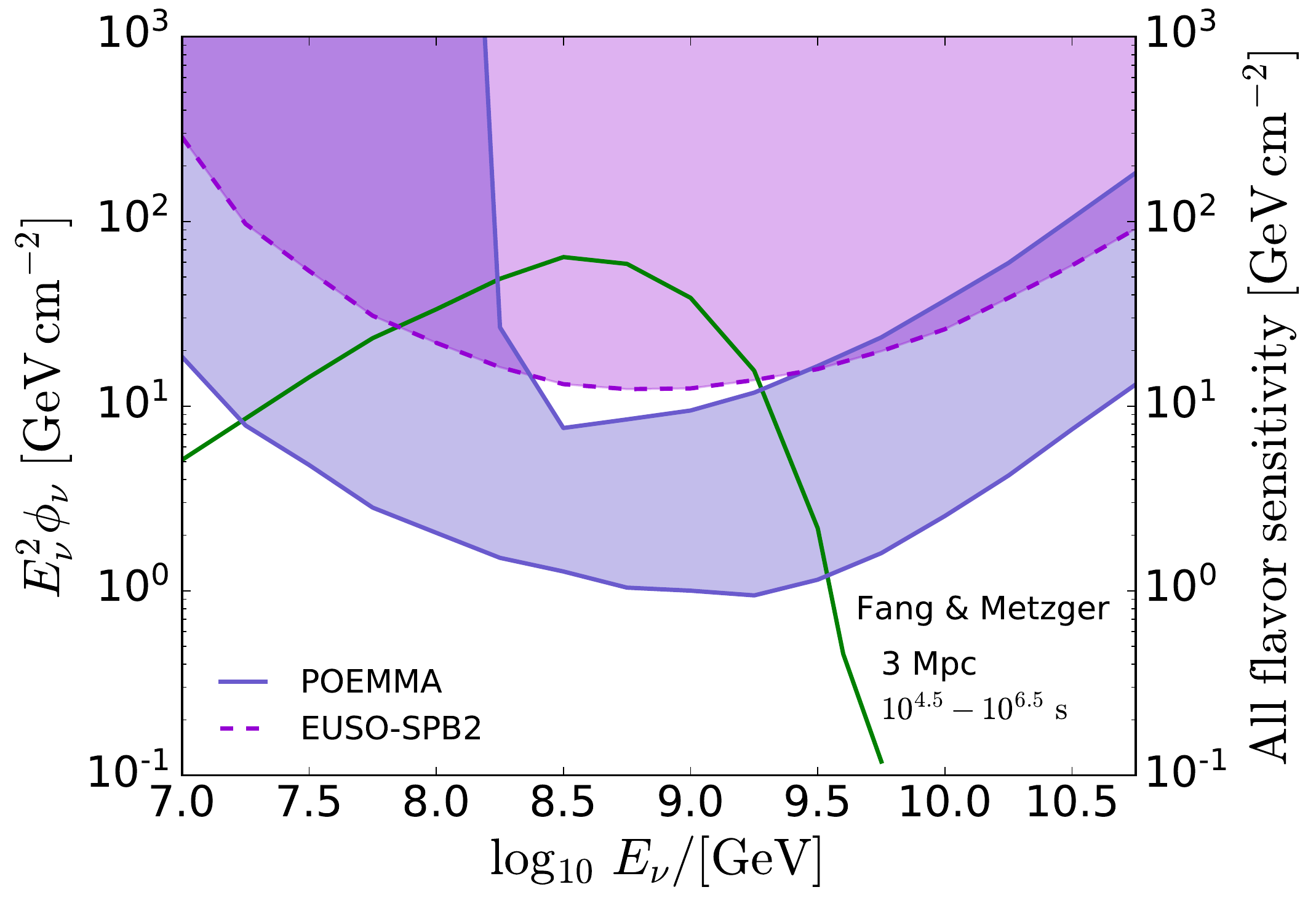}
       \hspace{1cm}
       \includegraphics[width=.48\textwidth]{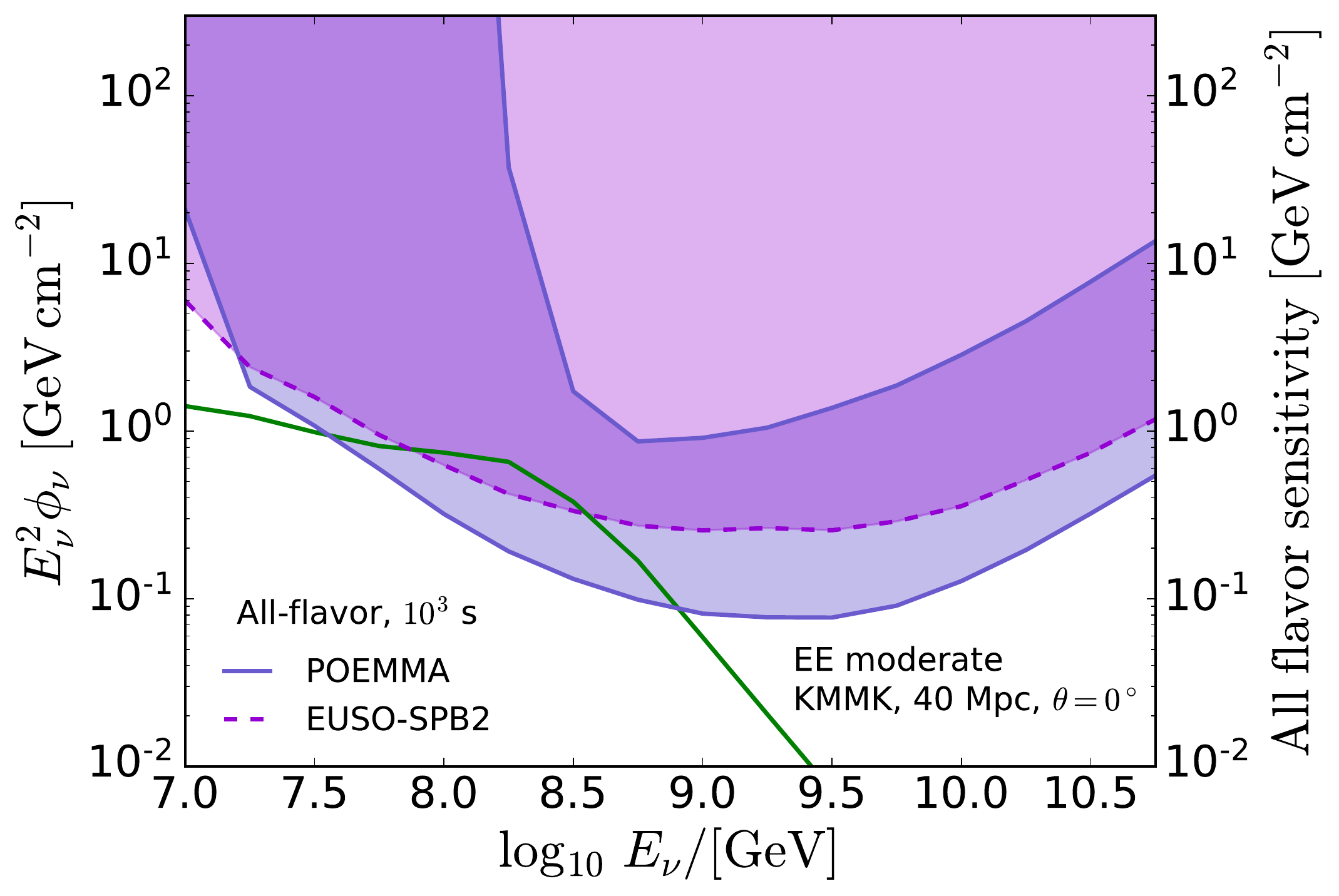}  
       \hspace{1cm}
       \caption{Long (left) and short (right) burst all-flavor neutrino fluence sensitivities for POEMMA (blue) and EUSO-SPB2 (violet) with parameters of Table 1 assuming instruments can slew over the $360^\circ$ in azimuth.
       }
       \label{fig:burst}
     \end{figure}

The sensitivity to neutrino point sources depends on the instantaneous effective area \cite{Venters:2019xwi}
 \begin{equation}
A_{\rm Ch} (s) \simeq \pi (v-s)^2 \times \left(\theta_{\rm Ch}^{\rm eff}\right)^2\ ,
\label{eq:ACh}
\end{equation}
the area subtended by the Cherenkov cone over the distance between the tau decay point and the telescope, $v-s$. For long neutrino bursts, the relevant quantity is the instantaneous effective area weighted by the observation probability, time averaged over the period $T_0$ of the satellite orbit (90 minutes for POEMMA) or the Earth's rotation of approximately one day (for EUSO-SPB2):
\begin{eqnarray}
A(\beta_{\rm tr}(t),E_{\nu},\theta,\phi)& \simeq& \int dP_{\rm obs}(E_{\nu},\beta_{\rm tr},s) A_{\rm Ch}(s) \\
\langle A(E_{\nu},\theta,\phi)\rangle_{T_0} &= &\frac{1}{T_0}\int _{t_0}^{t_0+T_0}\, dt A(\beta_{\rm tr}(t),E_{\nu},\theta,\phi)
\ .
\label{eq:avga}
\end{eqnarray}
The sensitivity is determined from 
\begin{equation}
\phi_{\rm sens} = \frac{2.44\times 3}{\ln (10)\times E_{\nu_\tau}\times \langle A
(E_{\nu},\theta,\phi)
\rangle_{T_0} \times f_t}\,,
\end{equation}
for the all-flavor fluence $\phi_\nu$.

The left panel of fig. \ref{fig:burst} shows the long neutrino burst sensitivities for POEMMA and EUSO-SPB2. For POEMMA, the sensitivity is evaluated assuming stereo viewing, and an overall duty cycle $f_t$ of 30\% is used to account for the time that the Sun and Moon prohibit observing. Dual viewing mode for POEMMA yields the best long neutrino burst sensitivity a factor $\sim 5$ higher (weaker) than in stereo mode for $10^7$ GeV. Stereo and dual viewing modes have very similar sensitivities above $E_\nu \sim 3\times 10^8$ GeV.
For EUSO-SPB2, the sensitivity to a long neutrino burst of 30 days $=10^{6.4}$ s comes from the 30 day average of the effective area, assuming a 20\% duty cycle because of the Sun and Moon. Since there are some regions of the sky inaccessible to EUSO-SPB2, we do not plot an upper bound to the sensitivity band. POEMMA has the advantage of full sky coverage over time.
We also show the prediction of Fang and Metzger's binary neutron star all-flavor neutrino fluence for a merger at 3 Mpc distance over a time interval of $10^{4.5}-10^{6.5}$ s \cite{Fang:2017tla}.

For short neutrino bursts, we assume that a $10^3$ s neutrino burst begins when the source is in an optimal position for observation by either instrument. The best short neutrino burst sensitivities are shown in the right panel of fig. \ref{fig:burst},  with the duty cycle  $f_t=1$, part of the definition of ``best.'' For a short neutrino burst, POEMMA does not have time to slew into stereo mode, so the sensitivity is determined in dual mode. Also shown in the right panel is the on-axis prediction for a short gamma ray burst at 40 Mpc with moderate extended emission predicted by Kimura, Murase, Meszaros and Kiuchi (KMMK) \cite{Kimura:2017kan}. The short neutrino burst sensitivity of EUSO-SPB2 is much closer to the POEMMA short neutrino burst sensitivity than for their respective long neutrino burst sensitivities, and even better than POEMMA at $E_\nu=10^7$ GeV. The low energy results for EUSO-SPB2 come from the advantage of being closer to the developing shower. There is boost in the effective area through $A_{\rm ch}$ with larger distances $(v-s)$ from the tau decay, at the cost of a lower photon density at the telescope. At low energies, the photon density threshold is the determining factor.

Both telescopes will require external (GCN) alerts for short and long neutrino bursts. With five years of viewing, POEMMA has greater potential for transient observations from tidal disruption events, binary black hole mergers and binary neutron star mergers \cite{Venters:2019xwi}. EUSO-SPB2 will be a pathfinder instrument. The sensitivities shown here indicate that sources in nearby galaxies are accessible for detection if those sources are located in optimal regions of the sky \cite{Venters}. 

\section{Modeling uncertainties}

The evaluations of the sensitivities to the diffuse neutrino flux and the neutrino point source fluence rely on modeling the tau neutrino and tau propagation in the Earth, the up-going tau air shower and its detection. In this section, we illustrate the impact of modeling uncertainties in tau propagation in the Earth (in $p_{\rm exit}$) and the shower modeling, which impacts $p_{\rm det}$. 

Neutrino propagation in the Earth has a primary uncertainty in the ultrahigh energy extrapolation of the neutrino cross section. Once a tau is produced, it propagates with electromagnetic energy loss dominated by photonuclear processes. Again, the high energy behavior of tau photonuclear scattering requires extrapolations. Here, we focus on electromagnetic energy loss. Two examples of parameterizations of the photonuclear energy loss are those of 
Abramowitz et al. (ALLM) \cite{allm} and Block et al. (BDHM) \cite{Block:2014kza}. The blue region between the solid black (ALLM) and blue (BDHM) curves shows an approximate theoretical uncertainty range for photonuclear energy loss. Both solid curves come from an Earth model based on the Preliminary Reference Earth Model \cite{Dziewonski:1981xy} with an outer layer of water of depth 3 km. The black dashed line shows the result when the final layer of Earth is rock. At low energies, most of the neutrino production of taus is in the final layer, so the rock density enhances the probability for the neutrino to interact. At high energies, tau can propagate farther. Propagation in water entails less energy loss than in rock, so the sensitivity is best over water at high energies \cite{PalomaresRuiz:2005xw}.
\begin{figure}
       \hskip 1.5in
       \includegraphics[width=.48\textwidth]{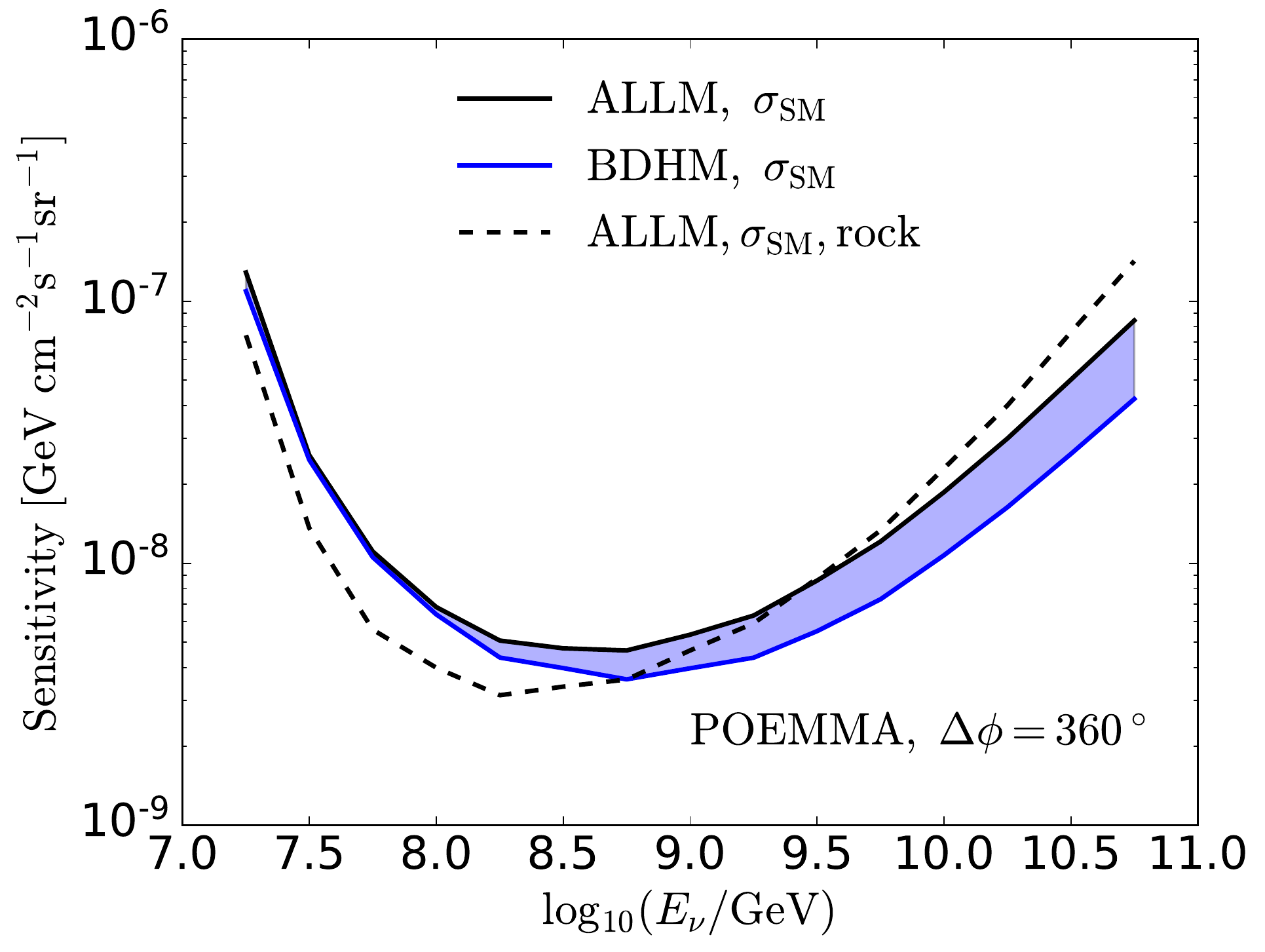}
       \caption{
       Comparison of POEMMA ($\Delta\phi=360^\circ$) sensitivity evaluations for the diffuse all-flavor neutrino flux  with ALLM and BDHM tau photonuclear energy loss extrapolations with 3 km water in the outer layer of the Earth (solid curves) and with a final layer of all rock (dashed). The figure is updated from ref. \cite{Reno:2019jtr}.
       }
       \label{fig:austin}
     \end{figure}

In refs. \cite{Reno:2019jtr} and \cite{Venters:2019xwi}, the photon density as a function of angle and altitude of the decay was determined using the Greisen parameterization of the EAS and a composite atmosphere model (see ref. \cite{Reno:2019jtr} for details) for $E_{\rm shr}=10^8$ GeV, then rescaled by a factor of $E_{\rm shr}/10^8$ GeV. In figs. \ref{fig:diffuse-limits} and \ref{fig:burst}, the shower energy is taken to be $0.5 E_\tau$. Recently, improvements to the air shower modeling have been reported in ref. \cite{Cummings:2020ycz}. There, the air shower modeling is based on CORSIKA, modified for upward showers. The longitudinal profile as a function of shower age shows longer shower tails than the Greisen parameterization, resulting in an increased effective aperture at high energies. 
Figure \ref{fig:austin} show a comparison of our results (black) for the sensitivity of the diffuse flux of POEMMA 
\cite{Reno:2019jtr} (left) and for EUSO-SPB2 (right)
with the results from the improved treatment of the showers (red) evident at high energies, illustrating the projected improved sensitivity at high energies for both instruments.
At low energies, the diffuse neutrino flux sensitivity of EUSO-SPB2 shows little change with the improved shower treatment. For POEMMA, the results are in qualitative agreement at low energies. The initial shower energy is expected to have a larger impact close to the sharper threshold of the higher altitude telescope.
The quantitative discrepancy between two results at low energy for POEMMA is under investigation. 

\begin{figure}
       \includegraphics[width=.48\textwidth]{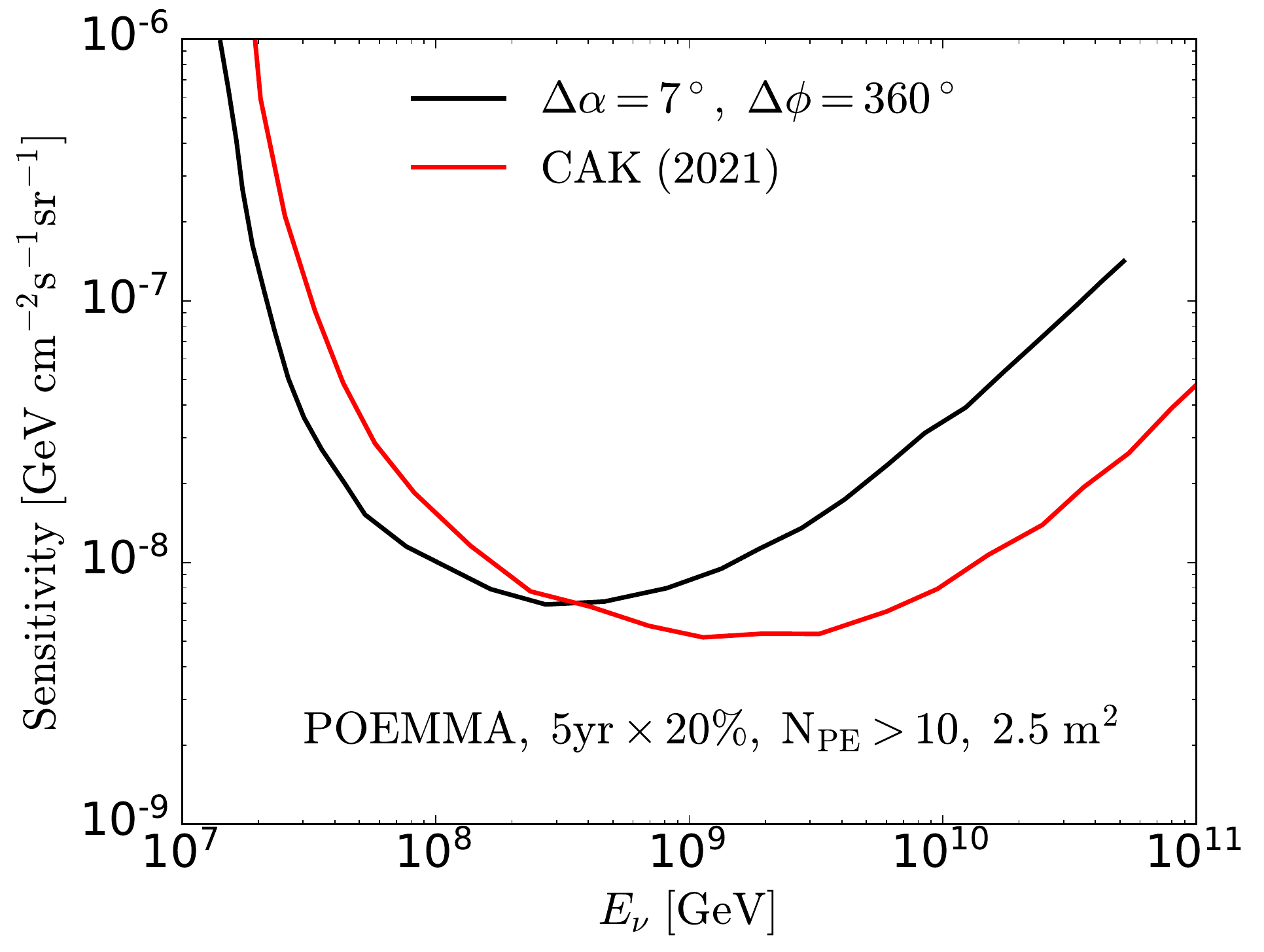}
       \includegraphics[width=.48\textwidth]{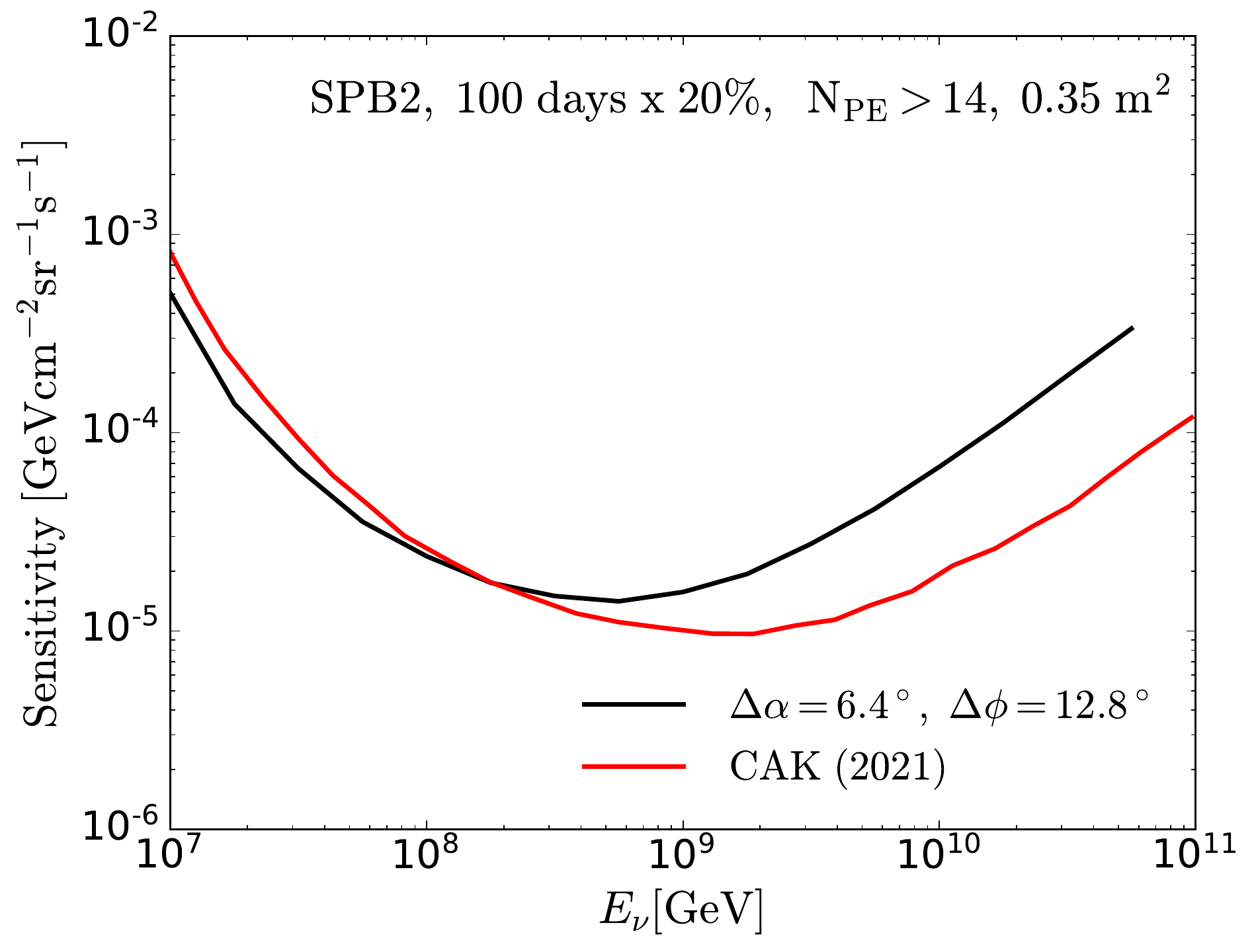}       
       \caption{
       Comparison of sensitivity evaluations for the all flavor diffuse neutrino flux $E_\nu^2\phi_{\nu+\bar{\nu}}$ following ref. \cite{Reno:2019jtr} in black and with the more detailed shower modeling of ref. \cite{Cummings:2020ycz} (black, labeled CAK (2021)).
       }
       \label{fig:austin}
     \end{figure}

Further studies of the impact of different models for the inputs to $P_{\rm obs}$ are in progress. The end-to-end simulation of neutrino to optical Cherenkov and geomagnetic radio signals from EAS in the {\texttt nuSpaceSim} package is under development \cite{Krizmanic:2020bdm}, with a component {\texttt nuPyProp} that evaluates the tau exit probabilities and energy distributions \cite{Patel}. 

\acknowledgments

This work is supported in part by NASA grants 80NSSC19K0626 at the University of Maryland, Baltimore County, 80NSSC19K0460 at the Colorado School of Mines, 80NSSC19K0484 at the University of Iowa, and 80NSSC19K0485 at the University of Utah, 80NSSC18K0464 at Lehman College, and under proposal 17-APRA17-0066 at NASA/GSFC and JPL, and by the US Department of Energy grant DE-SC-0010113. The conceptual design of POEMMA was supported by NASA Probe Mission Concept Study grant NNX17AJ82G for the 2020 Decadal Survey Planning. Contributors to this work were also supported in part by NASA awards 16-APROBES16-0023, NNX17AJ82G, NNX13AH54G, 80NSSC18K0246, and 80NSSC18K0473.



\clearpage
\section*{+POEMMA Collaboration}
%
%
\scriptsize
\noindent
A. V. Olinto,$^1$
J. Krizmanic,$^{2,3}$
J. H. Adams,$^4$
R. Aloisio,$^5$
L. A. Anchordoqui,$^6$
A. Anzalone,$^{7,8}$
M. Bagheri,$^9$
D. Barghini,$^{10}$
M. Battisti,$^{10}$
D. R. Bergman,$^{11}$
M. E. Bertaina,$^{10}$
P. F. Bertone,$^{12}$
F. Bisconti,$^{13}$
M. Bustamante,$^{14}$
F. Cafagna,$^{15}$
R. Caruso,$^{16,8}$
M. Casolino,$^{17,18}$
K. \v{C}erný,$^{19}$
M. J. Christl,$^{12}$
A. L. Cummings,$^{5}$
I. De Mitri,$^{5}$
R. Diesing,$^{1}$
R. Engel,$^{20}$
J. Eser,$^{1}$
K. Fang,$^{21}$
F. Fenu,$^{10}$
G. Filippatos,$^{22}$
E. Gazda,$^{9}$
C. Guepin,$^{23}$
A. Haungs,$^{20}$
E. A. Hays,$^{2}$
E. G. Judd,$^{24}$
P. Klimov,$^{25}$
V. Kungel,$^{22}$
E. Kuznetsov,$^{4}$
Š. Mackovjak,$^{26}$
D. Mandát,$^{27}$ L. Marcelli,$^{18}$ J. McEnery,$^{2}$ 
G. Medina-Tanco,$^{28}$ K.-D. Merenda,$^{22}$ S. S. Meyer,$^{1}$
J. W. Mitchell,$^{2}$ H. Miyamoto,$^{10}$ J. M. Nachtman,$^{29}$
A. Neronov,$^{30}$ F. Oikonomou,$^{31}$ Y. Onel,$^{29}$ 
G. Osteria,$^{32}$ A. N. Otte,$^{9}$ E. Parizot,$^{33}$ 
T. Paul,$^{6}$ M. Pech,$^{27}$ J. S. Perkins,$^{2 }$
P. Picozza,$^{18,34}$ L.W. Piotrowski,$^{35}$ 
Z. Plebaniak,$^{10}$ G. Prévôt,$^{33}$ P. Reardon,$^{4}$
M. H. Reno,$^{29}$ M. Ricci,$^{36}$ O. Romero Matamala,$^{9}$
F. Sarazin,$^{22}$ P. Schovánek,$^{27}$ V. Scotti,$^{32,37}$
K. Shinozaki,$^{38}$ J. F. Soriano,$^{6}$ F. Stecker,$^{2}$
Y. Takizawa,$^{17}$ R. Ulrich,$^{20}$ M. Unger,$^{20}$ T. M. Venters,$^{2}$ L. Wiencke,$^{22}$ D. Winn,$^{29}$ R. M. Young,$^{12}$
M. Zotov$^2$\\

%
\noindent
$^{1}$The University of Chicago, Chicago, IL, USA
$^{2}$NASA Goddard Space Flight Center, Greenbelt, MD, USA
$^{3}$Center for Space Science \& Technology, University of Maryland, Baltimore County, Baltimore, MD, USA
$^{4}$University of Alabama in Huntsville, Huntsville, AL, USA
$^{5}$Gran Sasso Science Institute, L’Aquila, Italy
$^{6}$City University of New York, Lehman College, NY, USA
$^{7}$Istituto Nazionale di Astrofisica INAF-IASF, Palermo, Italy
$^{8}$Istituto Nazionale di Fisica Nucleare, Catania, Italy
$^{9}$Georgia Institute of Technology, Atlanta, GA, USA
$^{10}$Universita’ di Torino, Torino, Italy
$^{11}$University of Utah, Salt Lake City, Utah, USA
$^{12}$NASA Marshall Space Flight Center, Huntsville, AL, USA
$^{13}$Istituto Nazionale di Fisica Nucleare, Turin, Italy
$^{14}$Niels Bohr Institute, University of Copenhagen, DK-2100 Copenhagen, Denmark
$^{15}$Istituto Nazionale di Fisica Nucleare, Bari, Italy
$^{16}$Universita’ di Catania, Catania Italy
$^{17}$RIKEN, Wako, Japan
$^{18}$Istituto Nazionale di Fisica Nucleare, Section of Roma Tor Vergata, Italy
$^{19}$Joint Laboratory of Optics,  Faculty of Science,  Palacký University,  Olomouc,  Czech Republic
$^{20}$Karlsruhe Institute of Technology, Karlsruhe, Germany
$^{21}$Kavli Institute for Particle Astrophysics and Cosmology, Stanford University, Stanford, CA94305, USA
$^{22}$Colorado School of Mines, Golden, CO, USA
$^{23}$Department of Astronomy, University of Maryland, College Park, MD, USA
$^{24}$Space Sciences Laboratory, University of California, Berkeley, CA, USA
$^{25}$Skobeltsyn Institute of Nuclear Physics, Lomonosov Moscow State University, Moscow,Russia
$^{26}$Institute of Experimental Physics, Slovak Academy of Sciences, Kosice, Slovakia
$^{27}$Institute of Physics of the Czech Academy of Sciences, Prague, Czech Republic
$^{28}$Instituto de Ciencias Nucleares, UNAM, CDMX, Mexico
$^{29}$University of Iowa, Iowa City, IA, USA
$^{30}$University of Geneva, Geneva, Switzerland
$^{31}$Institutt for fysikk, NTNU, Trondheim, Norway
$^{32}$Istituto Nazionale di Fisica Nucleare, Napoli, Italy
$^{33}$Université de Paris, CNRS, Astroparticule et Cosmologie, F-75013 Paris, France
$^{34}$Universita di Roma Tor Vergata, Italy
$^{35}$Faculty of Physics, University of Warsaw, Warsaw, Poland
$^{36}$Istituto Nazionale di Fisica Nucleare - Laboratori Nazionali di Frascati, Frascati, Italy
$^{37}$Universita’ di Napoli Federico II, Napoli, Italy
$^{38}$National Centre for Nuclear Research, Lodz, Poland.

\section*{+JEM-EUSO Collaboration}
%
%
\scriptsize
\noindent
G.~Abdellaoui$^{ah}$, 
S.~Abe$^{fq}$, 
J.H.~Adams Jr.$^{pd}$, 
D.~Allard$^{cb}$, 
G.~Alonso$^{md}$, 
L.~Anchordoqui$^{pe}$,
A.~Anzalone$^{eh,ed}$, 
E.~Arnone$^{ek,el}$,
K.~Asano$^{fe}$,
R.~Attallah$^{ac}$
% ...
\\

\noindent
% ...
$^{ac}$ LPR at Dept. of Physics, Faculty of Sciences, University Badji Mokhtar, Annaba, Algeria 
% ...
$^{ah}$ Telecom Lab., Faculty of Technology, University Abou Bekr Belkaid, Tlemcen, Algeria 
% ...
$^{cb}$ APC, Universit\'e de Paris, CNRS, AstroParticule et Cosmologie, F-75013 Paris, France
% ...
$^{ed}$ Istituto Nazionale di Fisica Nucleare - Sezione di Catania, Italy
% ...
$^{eh}$ INAF - Istituto di Astrofisica Spaziale e Fisica Cosmica di Palermo, Italy
% ...
$^{ek}$ Istituto Nazionale di Fisica Nucleare - Sezione di Torino, Italy
$^{el}$ Dipartimento di Fisica, Universit\`a di Torino, Italy
% ...
$^{fe}$ Institute for Cosmic Ray Research, University of Tokyo, Kashiwa, Japan 
% ...
$^{fq}$ Nihon University Chiyoda, Tokyo, Japan 
% ...
$^{md}$ Universidad Polit\'ecnia de madrid (UPM), Madrid, Spain
% ...
$^{pd}$ University of Alabama in Huntsville, Huntsville, AL, USA
$^{pe}$ Lehman College, City University of New York (CUNY), NY, USA
% ...
\end{document}